\documentclass[aps,preprint,prd,showpacs,nofootinbib]{revtex4}

\usepackage{amsmath,amssymb}
\usepackage{graphicx,subfigure}
\usepackage{color,multirow}
\usepackage[colorlinks,linkcolor=magenta,anchorcolor=cyan,citecolor=blue,plainpages=false]{hyperref}

\hypersetup{colorlinks=true,
    breaklinks=true,
    pdfstartview=Fit,
    linkcolor=blue,
    citecolor=blue,
    urlcolor=blue}

\def\lf{\left}
\def\rt{\right}

\def\be{\begin{equation}}
    \def\ee{\end{equation}}
\def\ba{\begin{eqnarray}}
    \def\ea{\end{eqnarray}}

\begin{document}
\title{Can the universe experience an AdS landscape since matter-radiation equality?}

    \author{Hao Wang$^{1,2} $\footnote{\href{wanghao187@mails.ucas.ac.cn}{wanghao187@mails.ucas.ac.cn}}}
    \author{Yun-Song Piao$^{1,2,3,4} $ \footnote{\href{yspiao@ucas.ac.cn}{yspiao@ucas.ac.cn}}}

    \affiliation{$^1$ School of Fundamental Physics and Mathematical
        Sciences, Hangzhou Institute for Advanced Study, UCAS, Hangzhou
        310024, China}

    \affiliation{$^2$ School of Physical Sciences, University of
        Chinese Academy of Sciences, Beijing 100049, China}

    \affiliation{$^3$ International Center for Theoretical Physics
        Asia-Pacific, Beijing/Hangzhou, China}

    \affiliation{$^4$ Institute of Theoretical Physics, Chinese
        Academy of Sciences, P.O. Box 2735, Beijing 100190, China}

    %   \date{}
    \begin{abstract}

Though an anti-de Sitter (AdS) vacuum, corresponding to a negative
cosmological constant (NCC), can be not responsible for the
acceleration of current universe, it might coexist with one
evolving positive dark energy component at low redshift, as well
as with early dark energy around the recombination to solve the
Hubble tension. In this paper, we investigate the scenario with
one AdS vacuum around the recombination and one at low redshift,
and from both current observational and theoretical perspectives
preliminarily explore the possibility that the universe
experienced a landscape with multiple AdS vacua since
matter-radiation equality.

    \end{abstract}

    \maketitle
    \tableofcontents
    %\newpage

\section{Introduction}

The concordant $\Lambda$CDM model is thought to describe the
evolution of our universe since matter-radiation equality, in
which the dark energy (DE) has always been the cosmological
constant (CC). However, recently using their DESI DR2 baryon
acoustic oscillation (BAO) data combined with Planck CMB and
supernova data the DESI cooperation has found that DE is evolving
at $\gtrsim 3\sigma$ C.L. \cite{DESI:2025zgx}, see also
\cite{Lodha:2025qbg}\footnote{It seems to prefer that the state
equation parameter $w$ of DE crossed $w=-1$ at low redshift from
$w<-1$, a so-called Quintom scenario, see a recent review
\cite{Cai:2025mas}. }. The DR2 BAO results are consistent with
those of DESI DR1 \cite{DESI:2024mwx,DESI:2024aqx,DESI:2024kob}.
The relevant issues have been also intensively investigated since
DESI DR1,
e.g.\cite{Luongo:2024fww,Cortes:2024lgw,Carloni:2024zpl,Colgain:2024xqj,Giare:2024smz,Wang:2024dka,Yang:2024kdo,Park:2024jns,Shlivko:2024llw,Dinda:2024kjf,Seto:2024cgo,Bhattacharya:2024hep,Roy:2024kni,Wang:2024hwd,Notari:2024rti,Heckman:2024apk,Gialamas:2024lyw,Orchard:2024bve,Colgain:2024ksa,Wang:2024sgo,Li:2024qso,Ye:2024ywg,Giare:2024gpk,Dinda:2024ktd,Jiang:2024viw,Alfano:2024jqn,Jiang:2024xnu,Sharma:2024mtq,Ghosh:2024kyd,Reboucas:2024smm,Pang:2024qyh,Wolf:2024eph,RoyChoudhury:2024wri,Arjona:2024dsr,Wolf:2024stt,Giare:2024ocw,Wang:2024tjd,Alestas:2024eic,Carloni:2024rrk,Bhattacharya:2024kxp,Specogna:2024euz,Li:2024qus,Ye:2024zpk,Pang:2024wul,Akthar:2024tua,Colgain:2024mtg,daCosta:2024grm,Chan-GyungPark:2025cri,Sabogal:2025mkp,Du:2025iow,Ferrari:2025egk,Jiang:2025ylr,Peng:2025nez,Jiang:2025hco,Feng:2025mlo,Hossain:2025grx,Chakraborty:2025syu,Borghetto:2025jrk,Pan:2025psn,Pang:2025lvh,Wang:2025ljj,Kessler:2025kju,Yang:2025mws,Wolf:2025jed,RoyChoudhury:2025dhe,Specogna:2025guo,Ye:2025ark,Cheng:2025lod,Ling:2025lmw,Wolf:2023uno}.
The impacts of pre-recombination early dark energy (EDE) for this
result has been showed in e.g.
Refs.\cite{Wang:2024dka,Wang:2024tjd} with DESI DR1 and
Refs.\cite{Pang:2025lvh,Poulin:2025nfb} with DESI DR2. How the
evolving DE is preferred has been also verified for different
parameterisations of the state equation of DE in
e.g.Ref.\cite{Wolf:2025jlc}, and what current data can (and cannot
yet) say about evolving DE also has been reviewed in recent
Ref.\cite{Giare:2025pzu}.

It is well-known that anti-de Sitter (AdS) vacuum is theoretically
important and well-motivated.
%It is widely thought that the AdS
%vacua are among the best understood quantum gravity backgrounds by
%virtue of \textsf{the AdS/CFT correspondence}
%\cite{Maldacena:1997re}.
Though it is possible that dS vacua might exist
\cite{Kachru:2003aw,Kallosh:2004yh}, the construction of such
vacua in the string landscape seems to be at issues, e.g. recent
swampland conjecture \cite{Ooguri:2006in,Obied:2018sgi}, see
e.g.\cite{Palti:2022edh} for recent reviews and also
\cite{Kallosh:2019axr,Kallosh:2018psh}. However, AdS vacua are
ubiquitous in the string theory. The inflation in AdS landscape
and its implication for primordial perturbations have been
explored in e.g.
\cite{Felder:2002jk,Linde:2006nw,Li:2019ipk,Lin:2022ygd,Piao:2005ag,Piao:2004hr}.

Though an AdS vacuum, corresponding to a negative cosmological
constant (NCC) can be not responsible for the acceleration of
current universe, its impacts on our observable universe have been
widespread concerned recently. In the early resolution of the
Hubble tension\footnote{A $\sim 5\sigma$ discrepancy between the
Hubble constant reported by Riess et.al using Cepheids-calibrated
supernovae \cite{Riess:2021jrx} and that showed by the Planck
collaboration based on $\Lambda$CDM model using CMB data
\cite{Planck:2018vyg}.} (see \cite{CosmoVerse:2025txj} for recent
reviews), the AdS vacuum around the recombination
\cite{Ye:2020btb,Ye:2020oix,Jiang:2021bab,Ye:2021iwa,Wang:2022jpo}
(AdS-EDE), can efficiently boost the contribution of EDE
\cite{Karwal:2016vyq,Poulin:2018cxd,Smith:2019ihp,Kaloper:2019lpl,Agrawal:2019lmo,Alexander:2019rsc,Lin:2019qug,Sakstein:2019fmf,Niedermann:2019olb,Braglia:2020bym}\footnote{See
e.g.\cite{Braglia:2020iik,Braglia:2020auw} for early modified
gravity,
\cite{Poulin:2023lkg,Vagnozzi:2023nrq,McDonough:2023qcu,Ye:2023zel,Gsponer:2023wpm,Chudaykin:2020acu,Chudaykin:2020igl,Hill:2021yec,LaPosta:2021pgm,Simon:2022adh,Efstathiou:2023fbn,DAmico:2020ods,Krishnan:2020obg,Nunes:2021ipq,Hill:2020osr,Ivanov:2020ril,Goldstein:2023gnw,Vagnozzi:2021gjh,Allali:2021azp,Alexander:2022own,Clark:2021hlo,Reeves:2022aoi,Yao:2023qve}
for confronting EDE with recent observations and \cite{Ye:2022afu}
for the impact on the search for primordial gravitational waves.},
and lead a bestfit Hubble constant $H_0\sim 73$km/s/Mpc. It is
possible that the NCC might exist for a long period and then
rapidly switch its sign at certain redshift ($\Lambda_s$CDM)
\cite{Akarsu:2019hmw,Akarsu:2021fol,Akarsu:2022typ,Akarsu:2023mfb,Paraskevas:2024ytz,Akarsu:2025gwi}\footnote{The
sign-switching CC at low redshift has been combined with
pre-recombination new physics in Ref.\cite{Toda:2024ncp}.}, which
also can prefer a higher $H_0$, see also recent
\cite{Escamilla:2025imi}. It is also possible that at low
redshift, one NCC and one evolving quintessence-like ($w\geqslant
-1$) DE component coexist, which has been hinted by recent DESI
BAO data \cite{Wang:2024hwd,Mukherjee:2025myk,Notari:2024rti}, see
\cite{Dutta:2018vmq,Visinelli:2019qqu,Ruchika:2020avj,DiValentino:2020naf,Calderon:2020hoc,Sen:2021wld,Malekjani:2023dky}
for earlier works with pre-DESI data. The NCC might also explain
the unexpected massive galaxies observed by JWST at redshifts
$z\gtrsim5$ \cite{Adil:2023ara,Menci:2024rbq}. These results seem
to be suggesting that there might exist more than one AdS periods
during the evolution of our universe since matter-radiation
equality.

In this paper, we will show that a multiple-AdS universe since
matter-radiation equality, in particular the $w_0w_a$CDM model
with low-redshift NCC and AdS-EDE, can be compatible with recent
observations (Planck18+DESI+PantheonPlus+SH0ES dataset).
%and a pre-recombination resolution of the Hubble tension, such as
%AdS-EDE, might be still necessary.
Then in light of the potential significance of the AdS vacua on
both high-redshift (early) and low-redshift universe, we explore
the possible models that the universe enjoyed a landscape with
multiple AdS vacua, in which AdS-EDE-like evolution can be
incorporated into low-redshift DE model with NCC.

%We conclude in Section.\ref{chp4}.

%\label{chpt2}

\section{Compatibility with recent observations}

\subsection{Data and Method}

To check the compatibility of post-recombination AdS landscape, we
use \textbf{Planck 2018 CMB} dataset (low-l and high-l TT, TE, EE
spectra, and reconstructed CMB lensing spectrum
\cite{Planck:2018vyg,Planck:2019nip,Planck:2018lbu}), \textbf{DESI
BAO} data\footnote{Recent DESI DR2 BAO results are consistent with
those of DESI DR1, here what we used are DR1.}, \textbf{Pantheon
Plus+SH0ES} data (consisting of 1701 light curves of 1550
spectroscopically confirmed Type Ia SN coming from 18 different
surveys \cite{Scolnic:2021amr}, using \textbf{SH0ES} Cepheid
distances as a calibrator of the SN1a magnitude).

Here, the $w_0w_a$CDM+NCC model corresponds to the model where one
NCC and one evolving positive DE component coexist, see
e.g.Refs.\cite{Wang:2024hwd,Notari:2024rti}. They are related by
$\Omega_m+\Omega_r+\Omega_x+\Omega_\Lambda=1$ with the density
factions of the matter $\Omega_m$, the radiation $\Omega_r$, the
evolving positive DE component $\Omega_x$ and the CC
$\Omega_\Lambda$. The density parameter $\Omega_x$ of evolving DE
satisfies $\Omega_\mathrm{DE}=\Omega_x+\Omega_L\simeq0.7$, with
the state equation of $\Omega_x$
\cite{Chevallier:2000qy,Linder:2002et} \be
w(z)=w_0+w_a\frac{z}{1+z}.\ee
%where $w_0$ and $w_a$ represents the
%state equation and its derivative $dw/dz$ with respect to the
%redshift $z$, respectively, at $z=0$.
However, it is noteworthy that in the $w_0w_a$CDM+NCC model, the
state equation of total DE $\Omega_\mathrm{DE}$ is not $w$ but
\cite{Adil:2023ara}
\begin{equation}
w_\mathrm{eff}(z)=\frac{\Omega_x(1+z)^{2+3w_0+3w_a}[w_0+(w_0+w_a)z]e^{-\frac{3w_az}{1+z}}-\Omega_{\Lambda}}{\Omega_x(1+z)^{3(1+w_0+w_a)}e^{-\frac{3w_az}{1+z}}+\Omega_{\Lambda}},\label{we}
\end{equation}

To see the impact of NCC around the recombination, we will
consider the $w_0w_a$CDM+NCC model with AdS-EDE\footnote{Here, by
convention we call EDE with NCC around the recombination
AdS-EDE.}, see e.g.\cite{Ye:2020btb,Jiang:2021bab,Wang:2022jpo},
in which a phenomenological EDE with an AdS vacuum started at the
redshift $z\sim3000$, while the post-recombination universe is
$w_0w_a$CDM-like with low-redshift NCC. The AdS-EDE potential is
\cite{Ye:2020btb}
\begin{equation}\label{Vede}
    V(\phi)=\left\{\begin{array}{ll}
V_{0}\left(\dfrac{\phi}{M_{{p}}}\right)^{4}-\Lambda_{\mathrm{ads}},&\quad \mathrm{for} \quad \dfrac{\phi}{M_{{p}}}<\left(\dfrac{\Lambda_{{ads}}}{V_{0}}\right)^{1 / 4} \\
    0, &\quad \mathrm{for} \quad \dfrac{\phi}{M_{{p}}}>\left(\dfrac{\Lambda_{ads}}{V_{0}}\right)^{1 / 4}
    \end{array}\right.
\end{equation}
where $\Lambda_{ads}$ is the AdS depth\footnote{It is fixed by
$\alpha_{ads}\equiv {\Lambda_\mathrm{ads}\over
\rho_m(z_c)+\rho_r(z_c)}=3.79\times10^{-4}$ as an effective
shortcut to avoid bad convergence of the chain
\cite{Ye:2020btb}.}.

We modified the MontePython-3.6 sampler
\cite{Audren:2012wb,Brinckmann:2018cvx} and CLASS codes
\cite{Lesgourgues:2011re,Blas:2011rf} to perform our MCMC
analysis, and adopt the Gelman-Rubin convergence criterion with a
threshold $R-1<0.01$. The flat priors of all parameters are listed
in Table.\ref{prior}.

%The parameters of our MCMC are \{$\omega_b$, $\omega_{cdm}$,
%$H_0$, $\ln10^{10}A_s$, $n_s$, $\tau_{reio}$, $w_0$, $w_a$,
%$\Omega_\Lambda$\} for $w_0w_a$CDM with low-redshift negative CC
%and \{$\omega_b$, $\omega_{cdm}$, $H_0$, $\ln10^{10}A_s$, $n_s$,
%$\tau_{reio}$, $w_0$, $w_a$, $\Omega_\Lambda$, $f_{ede}$, $z_c$\}
%(the corresponding flat priors in Table.\ref{prior}) for that with
%simultaneous low-redshift negative CC and AdS-EDE, respectively.

    \begin{table}[htbp]
    \centering
    \begin{tabular}{cc}
        \hline
        Parameters&Prior\\
        \hline
        $100\omega_b$&[None, None]\\
        $\omega_{cdm}$&[None, None]\\
        $H_0$&[65, 80]\\
        $\ln10^{10}A_s$&[None, None]\\
        $n_s$&[None,None]\\
        $\tau_{reio}$&[0.004, None]\\
        \hline
        $f_{ede}$&[$10^{-4}$, 0.3]\\
        $\ln(1+z_c)$&[7.5, 9.5]\\
        \hline
        $w_0$&[-2, 0.34]\\
        $w_a$&[-3, 2]\\
        $\Omega_\Lambda$&[-2, 0.6]\\
        \hline
    \end{tabular}
\caption{\label{prior} The priors of primary parameters we adopt
in MCMC analysis. Here, instead of \{$V_0$, $\phi_\text{i}$\} in
(\ref{Vede}) we choose \{$\ln(1+z_c)$, $f_{ede}$\} as the MCMC
parameters of EDE, see \cite{Ye:2020btb}, where $f_{ede}$ is the
energy fraction of EDE when the field starts to roll at the
redshift $z_c$. }
 \end{table}

\subsection{Results}

    \begin{table*}[htbp]
    \centering
    \begin{tabular}{c|c|c}
        \hline
        Parameters&$w_0w_a$CDM+AdS&$w_0w_a$CDM+AdS+NCC\\
        \hline
        $100\omega_b$&2.332(2.348)$\pm$0.020&2.333(2.339)$\pm$0.018\\
        $\omega_{cdm}$&0.135(0.132)$\pm$0.002&0.135(0.134)$\pm$0.002\\
        $H_0$&72.71(72.83)$\pm$0.68&70.63(73.52)$\pm$4.14\\
        $\ln10^{10}A_s$&3.0372(3.079)$\pm$0.015&3.068(3.054)$\pm$0.015\\
        $n_s$&0.995(0.999)$\pm$0.008&0.996(0.995)$\pm$0.005\\
        $\tau_{reio}$&0.054(0.054)$\pm$0.008&0.053(0.046)$\pm$0.008\\
        \hline
        $f_{ede}$&0.114(0.109)$\pm$0.007&0.114(0.107)$\pm$0.008\\
        $\ln(1+z_c)$&8.117(8.281)$\pm$0.074&8.178(8.176)$\pm$0.076\\
        \hline
        $w_0$&-0.861(-0.861)$\pm$0.056&-0.848(-0.887)$\pm$0.076\\
        $w_a$&-0.569(-0.566)$\pm$0.228&-0.563(-0.509)$\pm$0.314\\
        $\Omega_\Lambda$&-&0.071(-0.002)$\pm$0.122\\
        \hline
        $\Omega_m$&0.300(0.297)$\pm$0.005&0.299(0.291)$\pm$0.006\\
        $S_8$&0.863(0.855)$\pm$0.012&0.848(0.856)$\pm$0.036\\
        \hline
        $\chi^2_\mathrm{CMB}$&2777.17&2776.63\\
        $\chi^2_\mathrm{DESI}$&15.67&15.01\\
        $\chi^2_\mathrm{Pantheon+SH0ES}$&1288.12&1288.13\\
        \hline
        $\chi^2_\mathrm{tot}$&4080.98&4079.77\\
        \hline
    \end{tabular}
\caption{\label{MCtable}  \textbf{Mean (bestfit) values and
1$\sigma$ regions of the parameters of the $w_0w_a$CDM+CC+AdSEDE
model.} The dataset used is the Planck18+DESI+Pantheon Plus+SH0ES
dataset.}
\end{table*}

In light of our results in Ref.\cite{Wang:2024hwd}, though the
bestfit value of $H_0$ for the $w_0w_a$CDM+NCC model with the
SH0ES prior is larger than that without it, but it is not enough
to solve the Hubble tension. Thus a pre-recombination resolution
of the Hubble tension might be still necessary. It is interesting
to see what if our universe holds not only the NCC at low
redshift, but also the AdS-EDE at redshift $z\sim 1100$.

In Table.\ref{MCtable} we present our MCMC results. As in the
$w_0w_a$CDM+AdSEDE model, for the $\Omega_x$ component of DE,
$w_0=-1$ and $w_a=0$ is still included $\leq2\sigma$ C.L..
However, a NCC (the bestfit $\Omega_\Lambda\sim-0.002$) is only
slightly preferred by data. In particular, the constraint on $H_0$
is wider, which are 6 times larger than that of
$w_0w_a$CDM+AdSEDE. The fit is improved by $\Delta\chi^2\sim-1.2$,
while $\Delta\chi^2_\mathrm{CMB}\sim-0.5$. This indicates that
$w_0w_a$CDM with low-redshift NCC and AdS-EDE is not be ruled out
by recent data, thus multiple AdS vacua since matter-radiation
equality can be compatible with observations at present.

%and also accommodates $H_0\gtrsim 73$.

\begin{figure*}
   \includegraphics[width=1\columnwidth]{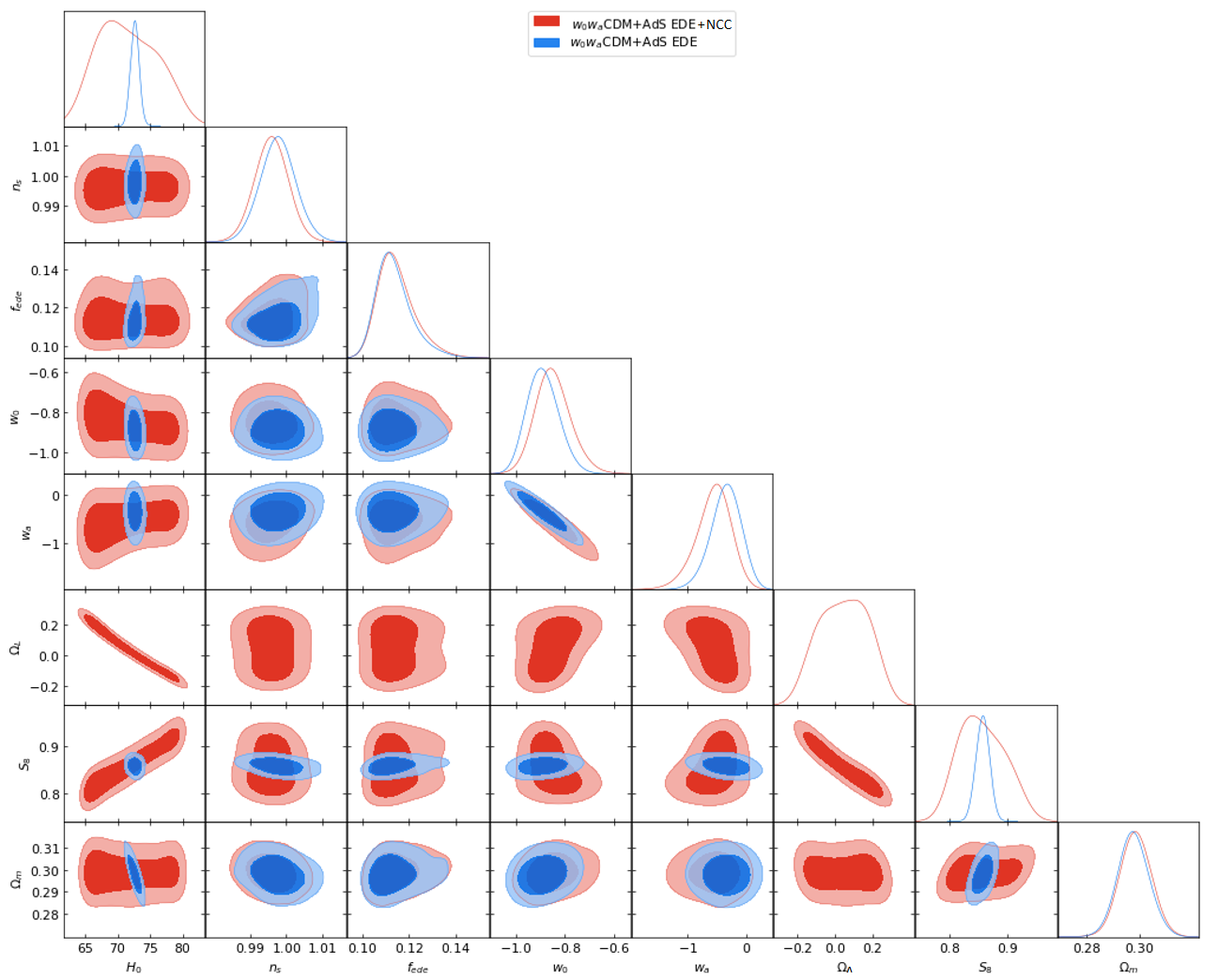}
    \caption{\label{MC} \textbf{2D contours of the
primary parameters at 68\% and 95\% CL for the
$w_0w_a$CDM+NCC+AdSEDE model}.}
\end{figure*}

In Fig.\ref{MC}, compared with the $w_0w_a$CDM+AdSEDE model, the
$2\sigma$ contour of $w_0-w_a$ in the $w_0w_a$CDM+AdSEDE+NCC
models slightly shifts down as a whole, but the cosmological
constant is still included $\leq2\sigma$ C.L. Here we find a
noteworthy degeneracy between $H_0$ and $\Omega_\Lambda$. The
negative correlation between $H_0$ and $\Omega_m$, which exists in
both $w_0w_a$CDM and $w_0w_a$CDM+AdSEDE models, is weakened as
well when $\Omega_\Lambda$ is considered. In particular, the
degeneracy between $\Omega_\Lambda$ and $w_0$ (also $w_a$) in
$w_0w_a$CDM+NCC (see \cite{Wang:2024hwd}) also becomes less
noticeable when AdS-EDE is concerned. This can be ascribed to an
adjustable sound horizon $r_s$ caused by EDE.
%The comoving distances observed by BAO are $D_M(z)/r_d$ and
%$D_H(z)/r_d$
%\begin{equation}\label{DMDH}
%    D_M(z)\equiv\int_{0}^{z}{cdz'\over H(z')},\quad D_H(z)\equiv {c\over
%        H(z)},
%\end{equation}
%where $r_d=\int_{z_d}^{\infty}{c_s(z)\over H(z)}$ is the sound
%horizon with $z_d\simeq1060$ at the baryon drag epoch and $c_s$
%the speed of sound.
Including the CC parameter $\Omega_\Lambda$ one obtains
\begin{equation}
    H(z)/H_0=\sqrt{\Omega_m(1+z)^3+\Omega_\Lambda+(1-\Omega_m-\Omega_\Lambda)(1+z)^{3(1+w_0+w_a)}e^{-3w_az/(1+z)}}.
    \label{HzH0}
\end{equation}
There is not additional parameters to adjust the sound horizon
$r_d=\int_{z_d}^{\infty}{c_s(z)dz\over H(z)}$ in the
$w_0w_a$CDM+NCC model ($z_d\simeq1060$ at the baryon drag epoch
and $c_s$ the speed of sound), which makes
$D_{M}(z)=\int_{0}^{z}{cdz'\over H(z')}$ and $D_H(z)= {c\over
        H(z)}$ almost constrained by data.
%At the redshift observed by BAO $z\sim1$, we can estimate by
%(\ref{HzH0})
%\begin{equation}
%\frac{D(z=1)}{D(z=1)|_{\Omega_\Lambda=0}}\sim
%\frac{H(z=1)}{H(z=1)|_{\Omega_\Lambda=0}}\sim0.2\Omega_\Lambda,\label{HL}
%\end{equation}
%and thus $\Omega_\Lambda$ have little impact on $H_0$ in the
%$w_0w_a$CDM+NCC model.
However, the inclusion of AdS-EDE alters $r_s^*$ ($r_s^*$ is
nearly equal to $r_d$), which makes $H_0$ and $\Omega_\Lambda$
have a larger range to fit the observations, while the bestfit
value is still $H_0\sim73$ km/s/Mpc.

%$H_0$ is larger than
%(\ref{HL}) estimated as $\Omega_\Lambda$ is expected to affect on
%$D_{M,H}(z)$ more significantly at high redshift.

\begin{figure*}
   \includegraphics[width=0.9\columnwidth]{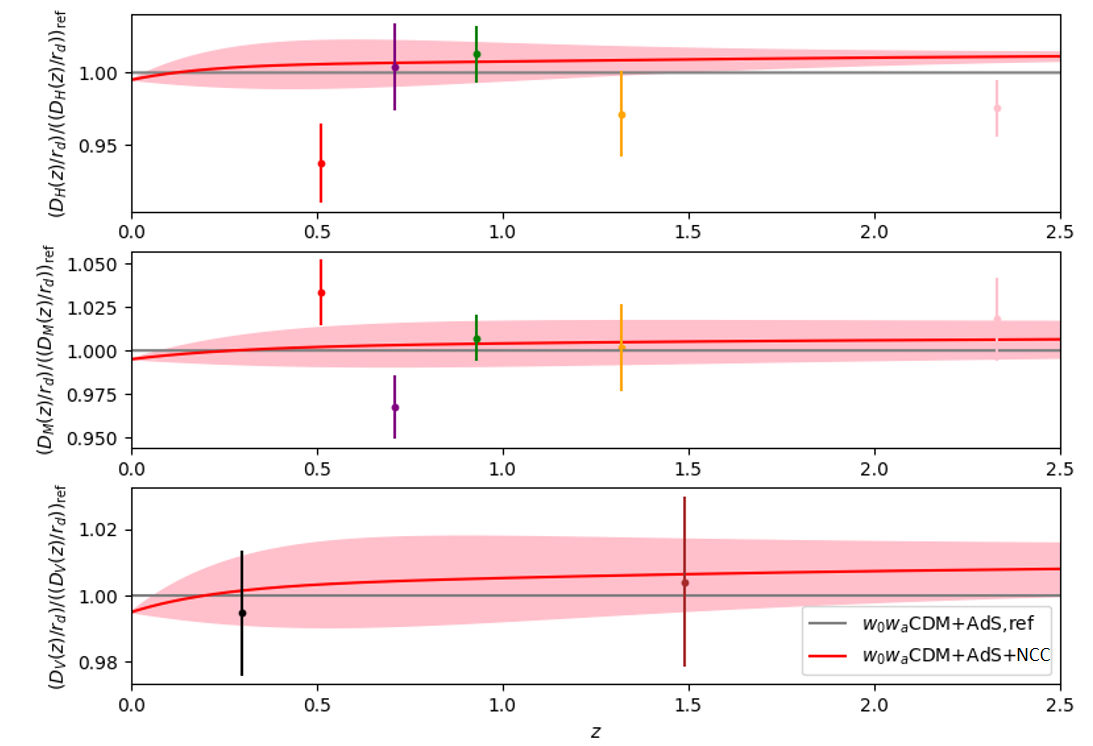}
    \caption{\textbf{Residuals of $D_H(z)/r_d$, $D_M(z)/r_d$ and $D_V(z)/r_d$
        for the bestfit values of the $w_0w_a$CDM+NCC+AdSEDE model with respect
        to the $w_0w_a$CDM+AdSEDE model}. The red
shadows are $1\sigma$ regions of $D_{H,M,V}(z)/r_d$,
respectively.} \label{DHDV}
\end{figure*}

In Fig.\ref{DHDV}, we plot the evolutions of $D_{H,M,V}(z)/r_d$
for the bestfit values of the $w_0w_a$CDM+NCC+AdSEDE model with
respect to the $w_0w_a$CDM+AdSEDE model, where $D_V/r_d$ is the
angle-averaged quantity with
$D_V(z)\equiv\left(zD_M(z)^2D_H(z)\right)^{1/3}$. The NCC
suppresses $D(z)/r_d$ at low redshift, keeps the fit to BAO, and
can accommodates a lower $r_s^*$ caused by AdS-EDE.

However, such a NCC did not affect the impact of AdS-EDE on the
primordial scalar spectral index $n_s$. It has been found that
$n_s=1$ ($n_s-1\sim {\cal O} (0.001)$) is preferred
\cite{Ye:2020btb,Ye:2021nej,Jiang:2022uyg,Smith:2022hwi,Jiang:2022qlj,Peng:2023bik}
when the pre-recombination EDE is considered, and see e.g.recent
Refs.\cite{Kallosh:2022ggf,Braglia:2020bym,Ye:2022efx,Jiang:2023bsz,Braglia:2022phb,DAmico:2021fhz,Giare:2023wzl,Giare:2024akf}
for its implications for inflation models, see also
\cite{DiValentino:2018zjj,Giare:2022rvg}. Here, we still have
$n_s=1$, ($n_s=0.996\pm0.005$), see Table.\ref{EDE}, however,
interestingly it seems $n_s=1$ for a wider $H_0$ and the scaling
relation $\delta n_s\simeq 0.4{\delta H_0\over H_0}$ in
Ref.\cite{Ye:2021nej} seems to be broken. It is noted that in
$w_0w_a$CDM+NCC model we have $n_s=0.967$ for $H_0=67.91$
\cite{Wang:2024hwd}, thus the issue relevant to $n_s$ is worth
further investigating.

\section{Modeling multiple AdS vacua since matter-radiation equality} \label{chap3}

\begin{figure}[htbp]
    \includegraphics[width=0.6\textwidth]{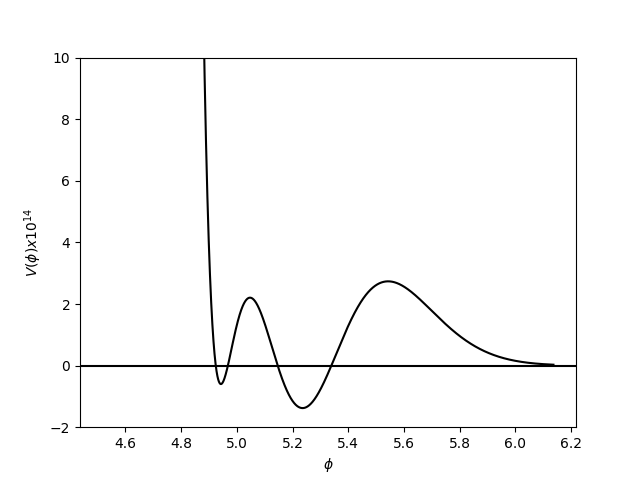}
\caption{\label{V1} \textbf{Potential with two AdS vacua.} The
first AdS minimum is at $\phi_0\simeq4.94$ and the second
$\phi_1\simeq5.24$, with $a_1=2\pi/100$, $a_2=2\pi/80$,
$A_1=-0.176$ and $A_2=0.98$. Here, we mark the first and second
maximum $V_1$ and $V_2$, respectively.}
\end{figure}

In this section, we will investigate the possibility of
integrating AdS-EDE and low-redshift NCC into one single
model\footnote{It has been argued that EDE can be implemented in
string theory \cite{McDonough:2022pku,Cicoli:2023qri}.}. It is
necessary to list the conditions that the models integrating
AdS-EDE and NCC at low redshift should satisfy. In the
corresponding model, initially the field is AdS-EDE-like with \be
V(\phi_i)\sim(0.1eV)^4,\ee \be |\partial^2_\phi
V(\phi_i)|\sim9H^2(z_c)\sim {(0.1eV)^4\over M_p^2}\label{EDE}\ee
must be required before the recombination ($z>1100$), see
e.g.Ref.\cite{Wang:2024jug}, however, at low redshift ($z\lesssim
5$) it is rolling towards an AdS vacuum and is quintessence-like
with \be V_\Lambda\sim(0.0001 eV)^4.\label{EDE2}\ee

Here, we focus on a string-motivated effective potential
\cite{Kallosh:2004yh}, see also
\cite{Bernardo:2020lar,Bernardo:2021vfw}\footnote{In string
theory, non-perturbative AdS vacua can be constructed, and it is
expected that at least one of them can be uplift to a dS vacuum.
Here, we do not consider the uplift to dS. }
\begin{equation}
    \begin{aligned}
V(\phi)=&{\lf(\sum_{i=1}^{N}
A_ia_i\exp(-a_ie^{\sqrt{2/3}\phi})\rt)^2}/{6
e^{\sqrt{2/3}\phi}}\\&+{\lf(\sum_{i=1}^{N} A_ia_i\exp(-a_i
e^{\sqrt{2/3}\phi})\rt)\lf(\sum_{i=1}^{N} A_i\exp(-a_i
e^{\sqrt{2/3}\phi})\rt)}/\lf({2 e^{2\sqrt{2/3}\phi}}\rt).
    \end{aligned}
    \label{Vphi}
\end{equation}
%\begin{equation}
%V(\sigma)=\left(\sum_{i=1}^N
%A_ia_ie^{-a_i\sigma}\right)^2/6\sigma+\left(\sum_{i=1}^N
%A_ia_ie^{-a_i\sigma}\right)\lf(\sum_{i=1}^N
%    A_ie^{-a_i\sigma}\rt)/2\sigma^2,
%\end{equation}
with free coefficients $A_i$ and $a_i$, where $N$ is the number of
non-perturbative contributions, see \cite{Bernardo:2020lar} for
details.
%According to Ref.\cite{Kallosh:2004yh}, the canonically
%normalized field is $\phi=\sqrt{3/2}\ln\sigma$, thus we have
It is interesting to note that this potential can have multiple
AdS vacua, depending on the different values of $A_i$ and $a_i$,
see Fig.\ref{V1}. Thus it might potentially serve as our model
that the universe at different redshift might experience different
AdS phases.

%In particular, at matter-radiation equality the field starts
%rolling from the leftmost slope and rolls over the first AdS at
%$z\sim 1000$ (AdS-EDE),

%\subsection{Three non-perturbative contributions}

In the case of $N=3$,
%the corresponding potential is
%\begin{equation}
%\begin{aligned} V(\phi)=&{\lf(\sum_{i=1}^{3}
%A_ia_i\exp(-a_ie^{\sqrt{2/3}\phi})\rt)^2}/{6
%e^{\sqrt{2/3}\phi}}\\&+{\lf(\sum_{i=1}^{3} A_ia_i\exp(-a_i
%e^{\sqrt{2/3}\phi})\rt)\lf(\sum_{i=1}^{3} A_i\exp(-a_i
%e^{\sqrt{2/3}\phi})\rt)}/\lf({2 e^{2\sqrt{2/3}\phi}}\rt).
%    \end{aligned}
%            \label{V3}
%\end{equation}
altering $A_{i}$ or $a_i$ ($i=1,2,3$) simultaneously does not
alter the shape of the potential, and thus only four of parameters
$A_{i}$ and $a_i$ ($i=1,2,3$) are significant. Here, we set
$A_3=-1.05$ and $a_3=2\pi/70$ for simplification. In this case, we
have only four free parameters \{$A_1, A_2, a_1, a_2$\}.

\begin{figure}[htbp]
    \includegraphics[width=0.6\textwidth]{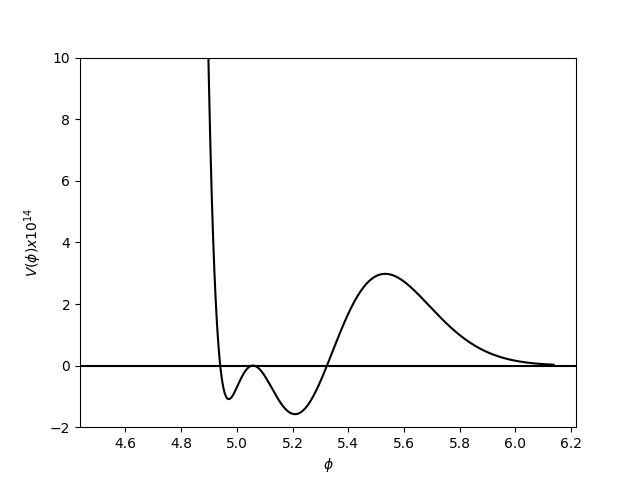}
\caption{\label{V2} \textbf{Potential in which AdS-EDE is
considered around the first AdS at $\phi_0$,} where
$\phi_0\simeq4.97$ with $A_1=-0.176757079$, $A_2=0.98015$,
$a_1=2\pi/100$ and $a_2=2\pi/80$. The first AdS minimum
$V(\phi_0)$ and the first peak $V_1$ satisfy
$|V(\phi_0)/V_1|\simeq10629.4$. }
  \end{figure}

\begin{figure}[htbp]
    \includegraphics[width=0.45\textwidth]{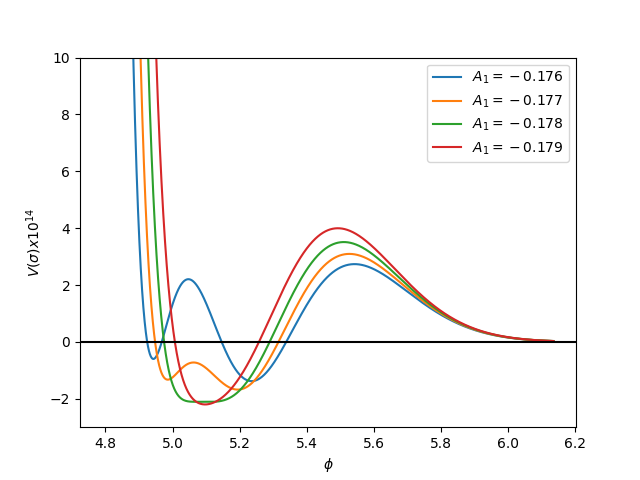}
    \includegraphics[width=0.45\textwidth]{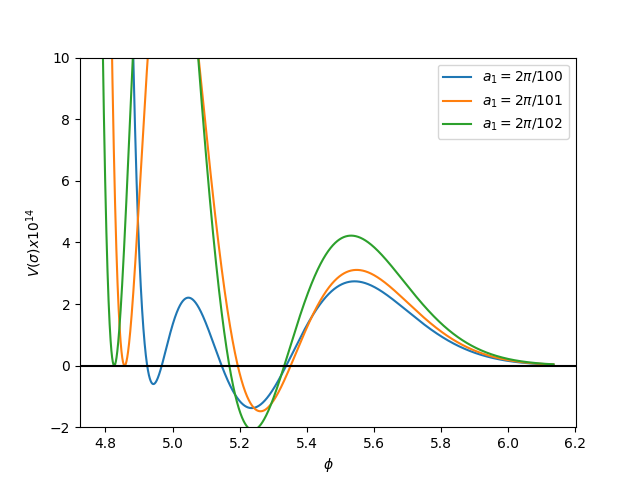}
    \includegraphics[width=0.45\textwidth]{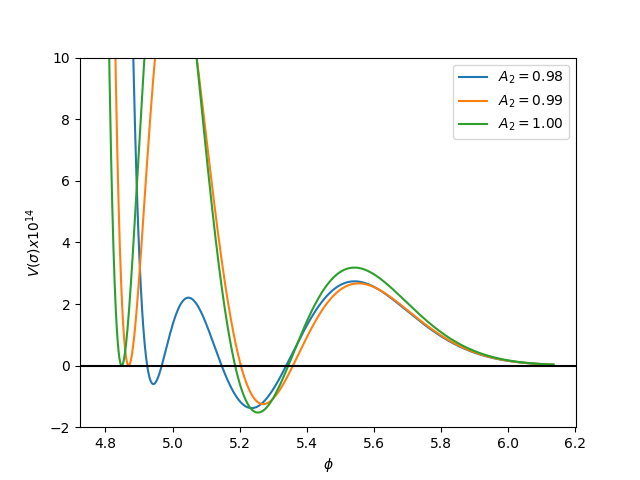}
    \includegraphics[width=0.45\textwidth]{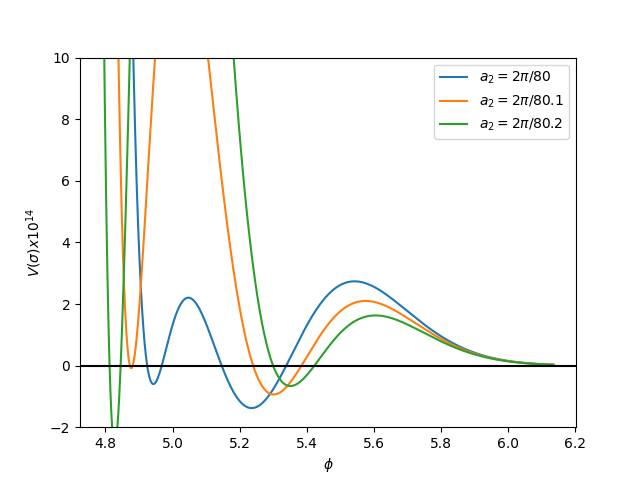}
\caption{\label{V5} \textbf{Potential with different $a_1$, $a_2$,
$A_1$ and $A_2$.} }
\end{figure}

In Fig.\ref{V2}, the first AdS minimum is responsible for AdS-EDE,
the field rolls over the first AdS minimum and eventually settles
on the first peak. It is also possible that for different values
of \{$A_1, A_2, a_1, a_2$\}, the second AdS minimum is responsible
for AdS-EDE, and the field rolls over the second AdS minimum and
eventually settles on the second peak.
%In the corresponding model, eventually the field might else stay
%near $V_2$, or overshoot $V_2$ and slow-roll finally,
%corresponding to the positive cosmological constant.
However, in both cases the compatibility with the condition
(\ref{EDE}) is hardly possible, see Fig.\ref{V5}.

%we have
%\begin{equation}
%    |V_{ads}/V_{\Lambda}|=\lf|{V(\phi_0)\over V_1}\rt|\sim10^4. \label{EDE1}
%\end{equation}

%Though (\ref{EDE1}) can be achieved simply by just tuning $A_1$
%and $A_2$, as illustrated in Fig.\ref{V2},

%The second AdS minimum has to satisfy
%  \begin{equation}
%    |V_{ads}/V_{\Lambda}|=\lf|{V(\phi_1)\over V_2}\rt|\sim10^4.\label{EDE2}
%  \end{equation}

%It might be also possible that the EDE rolls over both AdS vacua
%before settling on $V_2$. In this case, initially the EDE stays
%the leftmost slope of potential, and rolls over the first and
%second AdS at $z\sim 1000$, at least one of both AdS minima must
%satisfy
%\begin{equation}
%|V_{ads}/V_{\Lambda}|=\lf|{V(\phi_0)\over V_2}\rt|\quad or \quad
%\lf|{V(\phi_1)\over V_2}\rt|\sim10^4.\label{EDE2}
%\end{equation}
%In the corresponding model, a high middle hill is not conducive to
%EDE, thus both $V_1$ and $V_2$ has to be smaller than
%$|V(\phi_0)|$ or $V(\phi_1)$, which is difficult to be implemented
%with Eq.(\ref{V3}).

%Therefore, in the case with three non-perturbative contributions,
%it seems impossible to implement AdS-EDE.

%In Fig.\ref{V7}, it is showed that the second AdS $V(\phi_1)$ can
%be set to be smaller than $V_2$, while $V(\phi_0)$ is much larger.
%Thus it is possible that the field rolls over the first AdS, and
%stay for a while on middle hill at middle redshift, then it rolls
%over the second AdS ($z\sim 10$)and climbs the potential hill and
%like a dS-like DE at low redshifts ($z\sim {\cal O}(1)$).

%\begin{figure}[htbp]
%    \includegraphics[width=0.5\textwidth]{4.png}
%\caption{\label{V7}The EDE potential consists of two AdS minima,
%with $A_1=-0.169$, $A_2=0.969$, $a_1=2\pi/198.4$, $a_2=2\pi/160$
%and $a_3=2\pi/140$. $a_is$ have been shrunk together to ensure the
%excursion of the field is suitable and the shape is not altered.}
%\end{figure}

%\subsection{More non-perturbative contributions}

However, the case with $N=4$ is easier to satisfy (\ref{EDE}). In
the corresponding AdS-EDE model, for EDE around the first AdS
minimum, $|A_4a_4|\sim1$ and $a_4\gg1$ can lift the potential at
small $\phi$ but hardly change the second AdS minimum, while for
EDE in the second AdS minimum, a negative $A_4$ and $a_4\ll1$ can
magnify $|V(\phi_1)/V_2|$.

%Based on the result of Fig.\ref{V2}, we can add a term
%$A_4e^{-a_4\sigma}$ to satisfy $V(\phi_i)\sim(0.1eV)^4$.

\begin{figure}[htbp]
    \includegraphics[width=0.5\textwidth]{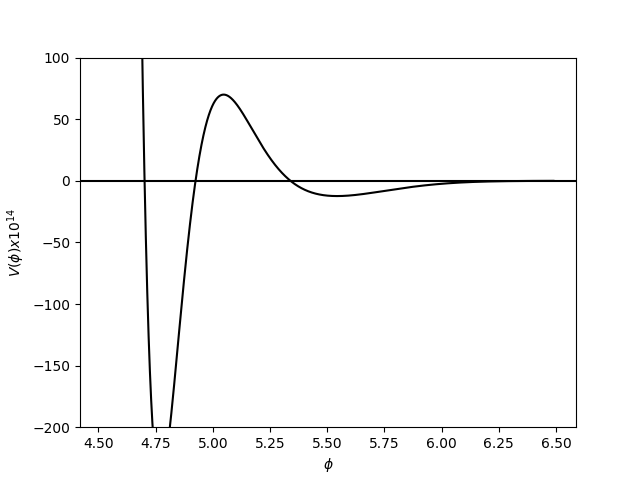}
\caption{\label{V6} \textbf{Potential in which AdS-EDE is
considered around the second AdS,} with $a_1=2\pi/100$,
$a_2=2\pi/80$, $a_4=2\pi/10000$, $A_1=-0.176$, $A_2=0.98$ and
$A_4=0.001$. $V(\phi_i)\sim(0.1eV)^4$ at $\phi_i\simeq5.13$ and
$\phi_1\simeq5.56$, where $|V(\phi_1)/V_2|\sim9978$.}
\end{figure}

In Fig.\ref{V2}, $V(\phi_i)\sim(0.1eV)^4$ suggests $\phi\sim4.89$,
where $|\partial^2_\phi V(\phi_i)|\sim100V(\phi_i)$. Considering
$a_4\gg1$, $|A_4|\ll1$ and $e^{-a_4\sigma}\ll1$, we get
\begin{equation}
    \begin{aligned}
\delta
V(\phi)=&A_4a_4e^{-a_4e^{\sqrt{2/3}\phi}}\\&\left[{{\sum_{i=1}^3
A_ia_i\exp(-a_ie^{\sqrt{2/3}\phi})}\over
{6e^{\sqrt{2/3}\phi}}}+{{\sum_{i=1}^3
A_i\exp(-a_ie^{\sqrt{2/3}\phi})}\over
{2e^{2\sqrt{2/3}\phi}}}\right],
    \end{aligned}\label{deltaV}
\end{equation}
\begin{equation}
    \begin{aligned}
\delta
V''(\phi)=A_4a_4^3e^{-a_4e^{\sqrt{2/3}\phi}}\left[\sum_{i=1}^3
A_i\exp(-a_ie^{\sqrt{2/3}\phi})\left(\frac{1}{2}+\frac{a_ie^{\sqrt{2/3}\phi}}{3}\right)\right].
    \end{aligned}
\end{equation}
The platform of quintessence-like DE is negligible compared to
AdS. The numerical result shows $a_4\sim2\pi/0.001$ to satisfy
(\ref{EDE}), while for a small $A_4$ and $a_4\ll1$ the potential
could satisfy (\ref{EDE}) and (\ref{EDE2}) simultaneously, see
Fig.\ref{V6}. Thus integrating AdS-EDE and low-redshift NCC into
one well-motivated model seems to be viable.

\begin{figure}[htbp]
   \includegraphics[width=0.45\textwidth]{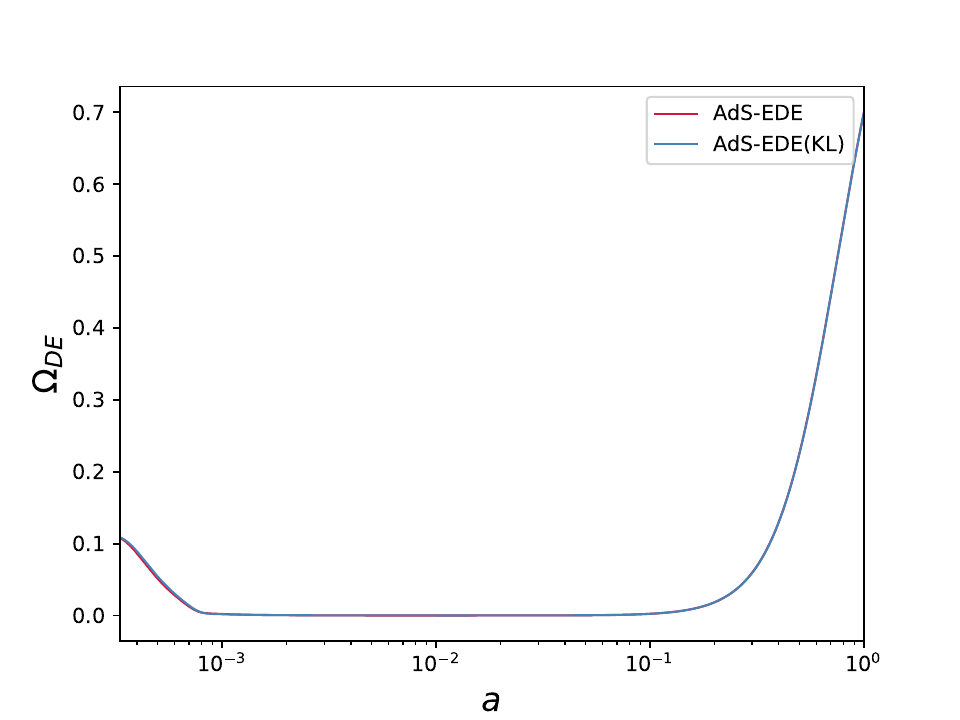}
   \includegraphics[width=0.45\textwidth]{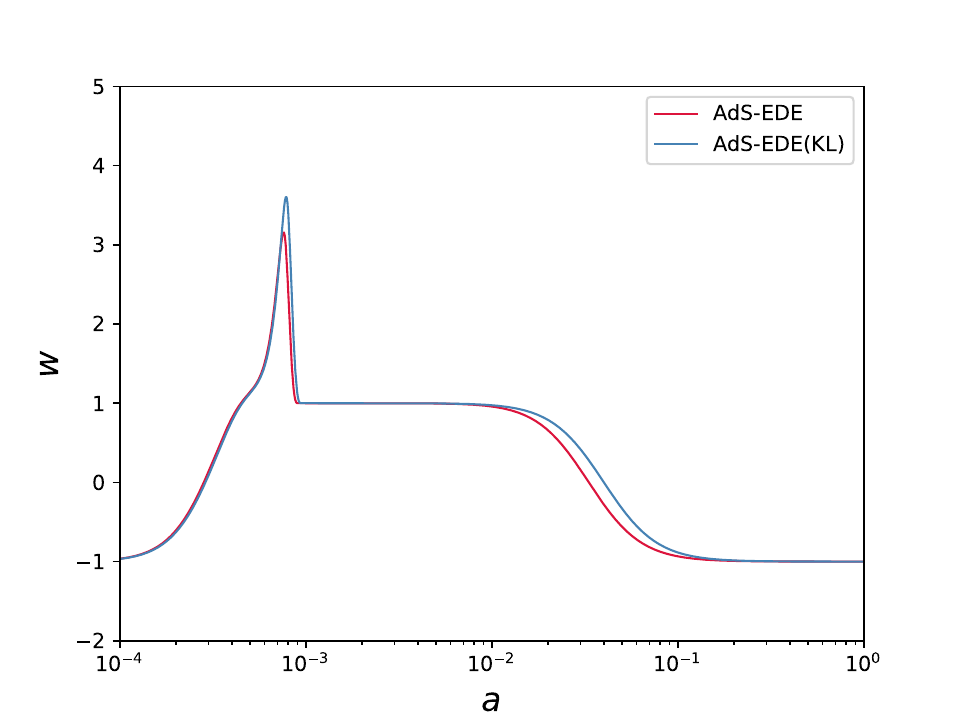}
\caption{\label{V61} \textbf{The evolutions of AdS-EDE fraction
$\Omega_{DE}$ and its state equation $w$.} The red line is that in
original AdS-EDE \cite{Ye:2020btb}, and the blue line is that for
the potential in Fig.\ref{V6}.}
\end{figure}

In Ref.\cite{Ye:2020btb}, the AdS-EDE model was build based on a
phenomenological potential (\ref{Vede}). As a coproduct, here with
well-motivated (\ref{Vphi}) we can present an AdS-EDE model in
Fig.\ref{V6}, in which the second AdS minimum is responsible to
implement AdS-EDE. In the corresponding model, initially the
scalar field could have similar evolution as the AdS-EDE in
Ref.\cite{Ye:2020btb}, see Fig.\ref{V61}.

%and then at low redshift it corresponds to quintessence-like DE
%with negative CC.

\section{Conclusion}\label{chp4}

In light of the theoretically significant values of the AdS vacua,
it is significant to explore how our universe can evolve in an AdS
landscape (with multiple AdS vacua) and its compatibility on the
observations. In this paper we investigate the possibility that
the universe experienced multiple AdS periods since the
matter-radiation equality, in particular the model which
integrates the pre-recombination AdS-EDE and the low-redshift NCC
($w_0w_a$CDM+AdSEDE+NCC).

Here, using Planck+DESI+PantheonPlus+SH0ES dataset we showed that
the $w_0w_a$CDM+AdSEDE+NCC model can be compatible with the
observations in which $H_0\sim 73$. Though compared with
$w_0w_a$CDM+AdSEDE, the fit is not improved significantly
($\Delta\chi^2\sim-1.2$ while
$\Delta\chi^2_\mathrm{CMB}\sim-0.5$), the NCC at low redshift can
not be ruled out. Thus it is possible that AdS-EDE and
low-redshift NCC can exist at different era of our universe since
matter-radiation equality\footnote{In $\Lambda_s$CDM model
\cite{Akarsu:2019hmw,Akarsu:2021fol,Akarsu:2022typ,Akarsu:2023mfb,Paraskevas:2024ytz,Akarsu:2025gwi},
there might also exist multiple sign-switchings of CC at different
redshifts.}.

It is interesting to comment the potential implications of our
results. The inclusion of AdS-EDE and low-redshift NCC can make
$w_0+w_a\geqslant -1$ (i.e. the evolving component of DE is the
quintessence) $<2\sigma$ consistent with recent DESI data, see
also \cite{Wang:2024dka,Wang:2024hwd}. In realistic model, the
scalar field can roll over different AdS vacua at different
redshift, around $z\sim 1100$ AdS-EDE is responsible for the
resolution to Hubble tension, while at $z\lesssim 10$ our current
dS-like era is only a momentary stage (it seems that the
quintessence-like DE is rolling towards an AdS vacuum, e.g.
earlier Refs.\cite{Kallosh:2002gf,Kallosh:2003mt,Cardenas:2002np},
and eventually our universe will collapse). The AdS vacua might
also get involved in inflation at very early universe, e.g.recent
multi-staged AdS inflation \cite{Li:2019ipk}, thus their relevant
phenomenology can be richer than expected. The NCC can also lead
to a more efficient growth of halo mass\footnote{The supermassive
primordial black holes also have similar effects for explaining
the high-redshift JWST observations, see
e.g.\cite{Huang:2024aog,Hai-LongHuang:2024vvz,Hai-LongHuang:2024gtx,Huang:2023chx,Huang:2023mwy}.},
thus it is also interesting to investigate the imprints of
multiple AdS stages on the high-redshift JWST observations, e.g.
recent Refs.\cite{Adil:2023ara,Menci:2024rbq}.

It should be noted that we only preliminarily explore the
possibility that the universe experiencing multiple AdS vacua
since the matter-radiation equality is compatible with recent
observations, as well as the possibility of implementing the
corresponding model in a string-motivated AdS landscape, however,
details relevant to the models and the compatibilities with
observations (e.g. recent ACT data \cite{ACT:2025fju}) still need
to further investigate. It is well known that recently the
understanding towards the black hole information paradox has made
a significant breakthrough by virtue of the AdS/CFT
correspondence, see \cite{Almheiri:2020cfm} for a recent review.
Thus in realistic world, if we have one or multiple AdS vacua, it
might be interesting to ask whether our universe can be
holographically encoded at AdS boundary or not, which might bring
us a novel and powerful insight into the origin of our universe.

Here we only show the possibility that the universe experience
multiple AdS vacua since matter-radiation equality, not whether it
is preferred or not. Perhaps, according to our results, recent
observations and the cosmological tensions is unveiling a corner
of a rich and colorful AdS landscape controlling the operation of
our universe.

%\begin{figure}[htbp]
%    \includegraphics[width=0.5\textwidth]{7.png}
%\caption{\label{V8}Illustration of an EDE potential with several
%AdS minima, where $a_1=2\pi/100$, $a_2=2\pi/80$, $a_4=2\pi/45$,
%$A_2=0.95$, $A_4=-0.013$ and $A_1=-0.1752A_1$.}
%\end{figure}

\section*{Acknowledgments}
This work is supported by NSFC, No.12475064, 12075246, National
Key Research and Development Program of China, No. 2021YFC2203004,
and the Fundamental Research Funds for the Central Universities.

\appendix

\end{document}